\documentstyle[prl,aps,multicol,epsf]{revtex}
\renewcommand{\narrowtext}{\begin{multicols}{2} \global\columnwidth20.5pc}
\renewcommand{\widetext}{\end{multicols} \global\columnwidth42.5pc} 

\begin{document}

\newcommand{\be}{\begin{equation}}
\newcommand{\ee}{\end{equation}}
\newcommand{\bea}{\begin{eqnarray}}
\newcommand{\eea}{\end{eqnarray}}
\newcommand{\nt}{\narrowtext}
\newcommand{\wt}{\widetext}


\title{Incipient nodal pairing 
in planar d-wave superconductors}

\author{D. V. Khveshchenko$^1$ and J. Paaske$^2$}

\address{$^1$ Department of Physics and Astronomy, University of North
Carolina, Chapel Hill, NC 27599\\ 
$^2$ Universit\"{a}t Karlsruhe, 
Institut f\"{u}r Theorie der Kondensierten Materie, Engesserstr. 7, 76128
Karlsruhe, Germany}
\maketitle

\begin{abstract}
The possibility of a second pairing transition $d\to d+is$ ($d+id^\prime$) in planar $d$-wave 
superconductors which occurs in the absence of external 
magnetic field, magnetic impurities or boundaries
is established in the framework of the non-perturbative phenomenon of dynamical chiral symmetry 
breaking in the system of $2+1$-dimensional Dirac-like nodal quasiparticles.  
We determine the critical exponents and quasiparticle spectral functions that characterize
the corresponding quantum critical behavior and discuss some of its potentially observable 
spectral and transport features.
\end{abstract}
\nt
The recent ARPES \cite{Valla} and optical conductivity  
\cite{Corson} studies of the superconducting high-$T_c$ compound 
$Bi_2Sr_2CaCu_2O_{8+\delta}$ have
challenged the physical picture of what was once thought to be 
an essentially mundane BCS-like pairing state with only weakly interacting
quasiparticles near the nodes of the $d_{x^2-y^2}$ order parameter.
Most notably, in a striking contrast with the well-known theoretical
prediction of the cubic in temperature quasiparticle damping rate obtained in the
conventional $d$-wave picture \cite{Scalapino}, both of the above probes \cite{Valla,Corson}
revealed a much weaker, approximately linear, temperature dependence.

Whether the above findings indicated a behavior
generic for all cuprates or not, their observation has prompted the idea of 
a quantum-critical behavior associated with a zero temperature quantum 
phase transition.
Should such a transition occur in a sufficient proximity to the 
physical slice (the standard "temperature-doping" plane) of the, 
conceivably, multi-dimensional phase diagram of the cuprates, 
the fluctuations of a new order parameter 
could provide for a strong enhancement of quasiparticle scattering 
resulting in the sought-after linear quasiparticle damping rate.

Recently, M. Vojta, Y. Zang, and S. Sachdev proposed 
a phenomenological theory of the nodal quasiparticles coupled to the fluctuations of a 
secondary ($is$ or $id_{xy}$) pairing order parameter. By applying a one-loop
$\epsilon$-expansion about $D=3+1$ space-time dimensions, they   
found an evidence of a quantum critical point residing 
inside the superconducting phase \cite{Sachdev}. 

In the present Letter, we develop a direct $2+1$-dimensional approach that 
will allow us to establish the existence of the critical point in question 
beyond perturbation theory and make quantitative predictions for     
the corresponding quantum critical behavior.

In order to facilitate the following discussion 
we make use of the rotationally-covariant Nambu representation where 
the $d$-wave quasiparticles are described in terms of the two species $(i=1,2)$, 
each having $N=2$ spin components labeled by $\alpha$, of the two-component 
Nambu spinors $\psi_{i\alpha}=(c_{\alpha}({\bf k}_i),\epsilon_{\alpha\beta}
c^{\dagger}_\beta(-{\bf k}_i))$ which are constructed from the quasiparticle states near  
the pairs of the opposite nodes located at $\pm{\bf k}_i=\pm (k_F,\pm k_F)/{\sqrt 2}$. 

Upon rotating the axes of the coordinate system by $\pi/4$,
we cast the linear in momentum kinetic energy of the nodal quasiparticles in the form of 
the (anisotropic) Dirac Lagrangian
\bea 
{\cal L}_\psi=i\sum^{N}_{\alpha=1}
\overline{\psi}_{1\alpha}(\gamma_0\partial_0-v_F\gamma_{1}\partial_{x}-
v_\Delta\gamma_2\partial_y)\psi_{1\alpha}\nonumber\\
+
(\psi_{1\alpha}\leftrightarrow \psi_{2\alpha}, ~~~
x\leftrightarrow y)
\label{Lf}
\eea    
where $v_F$ and $v_\Delta$ are related to the momentum 
derivatives of the normal state quasiparticle 
dispersion and the parent $d_{x^2-y^2}$-symmetrical gap function, respectively.
In Eq.(1), we use the irreducible $2\times 2$ representation 
for the $\gamma$-matrices in $2+1$ dimensions 
$\gamma_\mu=(\sigma_2, i\sigma_1, i\sigma_3)$ and the notation  
$\overline{\psi}_i=\psi^\dagger_i\gamma_0$. 
Also, in order to make Eq.(1) more symmetrical we perform a rotation in the Nambu space 
on one of the two fermion species: $\psi_{2\alpha}\to 
{\tilde \psi}_{2\alpha}=(\sigma_1+\sigma_3)\psi_{2\alpha}/{\sqrt 2}$.

Next, we combine the spinors $\psi_{1\alpha}$ and 
${\tilde \psi}_{2\alpha}$ into one four-component Dirac fermion 
$\Psi_\alpha=(\psi_{1 \alpha}, {\tilde \psi}_{2\alpha})$ and 
introduce a (reducible) 4-dimensional representation for the $\gamma$-matrices  
${\Gamma}_{\mu}=\gamma_{\mu}\otimes\sigma_3$ where the second factor in the tensor product
acts in the subspace of the two species.

In the phenomenological secondary pairing scenario, 
the nodal Dirac fermions couple to a real boson field $\phi$ corresponding to the imaginary
part of the total gap function. Its own critical dynamics is governed by a generic $\phi^4$-theory   
\be 
{\cal L}_{\phi}=
{1\over 2}({1\over c^2}(\partial_{0}\phi)^2-({\bf \nabla}\phi)^2-m^2\phi^2)
-{\lambda\over 24}\phi^4
\label{Lb} 
\ee 
At $T=0$ the theory (\ref{Lb}) undergoes the conventional $D=3$ 
Ising transition driven by the quartic self-interaction of the 
field $\phi$
that results in spontaneous breaking of the reflection ($Z_2$) symmetry $\phi\to -\phi$
as the parameter $m^2$ is tuned into its critical value $m^2_c$. 

Alternatively, the $Z_2$ symmetry breaking 
can also be driven by a linear coupling between $\phi$ and the nodal fermions
\be
{\cal L}_g=g\phi\sum^{N}_{\alpha=1}
\overline{\Psi}_{\alpha}{\bf \Lambda}\Psi_{\alpha}
\label{Lg}
\ee 
provided that both (\ref{Lf}) and (\ref{Lg})   
remain invariant under the discrete chiral transformation
\be
\phi\to -\phi,~~~~~~\Psi_{\alpha}\to\Gamma_5 \Psi_{\alpha},
~~~~~~{\overline \Psi}_{\alpha}\to -{\overline\Psi}_{\alpha}\Gamma_5
\label{Z2}
\ee
where $\Gamma_5=-{\bf 1}\otimes\sigma_2$ anticommutes with all $\Gamma_\mu$.

By choosing ${\bf \Lambda}={\bf 1}\otimes{\bf 1}$ 
in (\ref{Lg}), one couples $\phi$ to a $Z_2$-(and, accordingly, time reversal-) odd   
fermion mass operator. In terms of the original nodal quasiparticles,  
the latter reads as $\psi^\dagger_1\sigma_2\psi_1+\psi_2^\dagger\sigma_2\psi_2$
which appears to coincide with the order parameter of 
the $is$ pairing between the opposite nodes, while the chiral reflection (4) corresponds to a 
permutation of the two fermion species: $\psi_{1\alpha}\leftrightarrow \psi_{2\alpha}$.

With all the three velocities in (1) and (2) set 
equal ($v_F=v_\Delta=c=1$ ) the theory described by the Lagrangian 
${\cal L}_\psi+{\cal L}_\phi+{\cal L}_g$ becomes manifestly Lorentz- 
and $Z_2$-invariant (we will return to the spatially anisotropic case later). 

In fact, the above Lagrangian can be readily recognized as that of 
the Higgs-Yukawa (HY) model where the phenomenon of 
spontaneous chiral symmetry breaking (CSB) has long been discussed
as an alternative to the Higgs mechanism \cite{HY}.
Unlike its $1+1$- and $3+1$-dimensional counterparts which appear to be
asymptotically free and trivial (Gaussian), respectively, the 
$2+1$ dimensional HY model possesses a non-Gaussian infrared
fixed point at a finite coupling $g_c$ where CSB occurs 
via spontaneous generation of a non-zero fermion mass $M=g\langle\phi\rangle$. 

The above correspondence allows one to identify the CSB phenomenon with 
the conjectured second pairing transition $(d\to d+is)$ 
below which the nodal quasiparticles become fully gapped.
Albeit being absent in any finite order of perturbation theory, spontaneous CSB occurs
at the level of the non-perturbative mean field "gap equation" for the fermion mass 
\be
1={4Ng^2\over m^2}
T\sum_{\omega_n} \int {d^2{\bf k}\over (2\pi)^2}{1\over 
{\omega_n^2+{\bf k}^2+M^2(g,T)}} 
\label{gapeq}
\ee
which accounts for the tadpole ("Hartree") contribution to the fermion propagator \cite{HY}.

At $T=0$ Eq.(\ref{gapeq}) yields a critical coupling $g^2_c=\pi m^2/\Omega N$
as a function of the ultraviolet cutoff $\Omega$ set by 
the amplitude of the parent $d$-wave gap.
At strong coupling $(g>g_c)$, the chiral symmetry (4) 
is spontaneously broken, and the fermion mass scales alongside with the order 
parameter $M(g, 0)\propto <\phi>\propto  (g-g_c)^\beta$.

At finite $T$ the chiral symmetry gets restored above a transition 
line in the $g-T$ plane which terminates at the quantum 
critical point. In the symmetry broken (strong coupling, low-$T$) 
phase, the mean field equation (\ref{gapeq}) yields the fermion mass    
$M(g,T)=M(g,0)+ 2 T \ln[(1+{\sqrt{ 1-4\exp(-M(g,0)/T)}} )/2]
$
which vanishes along a critical line, resulting in the non-BCS-like relation  
$M(g, 0)/T^*_c=2\ln 2$ between the maximum gap and critical temperature $T_c^*$ 
of the nodal pairing. 

The quantum critical point associated with the CSB transition manifests itself via 
anomalous operator dimensions that can be found from the solution of 
the coupled Dyson equations for the fermion  
propagator $\langle\Psi{\overline \Psi}\rangle =
( Z_p {\not{\!p}} +\Delta_p)^{-1}$  
with all the tadpoles absorbed into the definition of the 
bare fermion mass (hereafter, $\not{\!p} = \Gamma_\mu p_\mu$ and $p^2=p^\mu p_\mu$): 
\bea
(Z_{p}-1)\not{\!p}+\Delta_{p}&=&\nonumber\\
&&\hspace*{-22mm}\int\!{d^3{k}\over (2\pi)^3}
\frac{Z_{p-k}(\not{\!p}-\not{\!k})+\Delta_{p-k}}{Z^{2}_{p-k}(p-k)^{2}+\Delta_{p-k}^2}
\frac{\Lambda_{p,k}^2}{k^{2}+m^{2}+\Pi_k},
\label{G}
\eea
and the fermion polarization operator 
\be
\Pi_k\!= N {\rm Tr}\!\!\int\!\!\!{d^3{p}\over (2\pi)^3}\Lambda_{p,k}^2
\frac{Z_{p}\not{\!p}+\Delta_{p}}{Z^{2}_pp^{2}+\Delta_{p}^2}
\frac{Z_{k+p}(\not{\!k}+\not{\!p})+\Delta_{k+p}}{Z^{2}_{k+p}(k+p)^{2}+
\Delta_{k+p}^2} 
\label{Pi}
\ee
which modifies the boson propagator $\langle\phi{\phi}\rangle =
(k^2+m^2+\Pi_k)^{-1}$.
In particular, the (ultraviolet divergent) 
momentum-independent part of (\ref{Pi}) contributes to the renormalized  
boson mass $\delta m^2=m^2+\Pi_0$ which vanishes at the critical point
$g=g_c$, consistent with Eq.(\ref{gapeq}).
 
In the lowest $1/N$ order, $Z_p=1, ~~~\Delta_p=M,~~~\Lambda_{p,k}=g$,
and the polarization operator (\ref{Pi}) assumes the form 
$$
\Pi_k=\Pi_0+{g^2N\over \pi}
\left(
{4M^2-k^2\over 2\sqrt{-k^2}}
\tan^{-1}\left({\sqrt{-k^2}\over 2M}\right)
-M\right)
$$
In the critical regime ($\delta m=M=0$), the infrared 
behavior of the boson propagator is governed solely by the fermion polarization
$\Pi_k-\Pi_0\propto{\sqrt {-k^2}}$ (hereafter, $\propto k$), 
while the bare $k^2$ term can be completely neglected.
This implies that at the CSB fixed point the mean field dimension of 
$\phi$ changes from its bare value $1/2$ to $1$,
thereby rendering the kinetic and quartic terms in Eq.(\ref{Lb}) totally irrelevant
in the renormalization group sense. 

We mention, in passing, that besides indicating
the existence of a fixed point at a critical coupling
$g^2_c\propto\epsilon=4-D$, the one-loop $\epsilon$-expansion 
used in Ref.\cite{Sachdev} also predicts
that the (formally relevant for any $\epsilon>0$) quartic term in (\ref{Lb}) scales 
towards strong coupling as well: $\lambda\to \lambda_c\propto\epsilon$,
which appears to be an artifact of the corresponding 
one-loop renormalization group equations.

With the only relevant term $m^2\phi^2$ in Eq.(\ref{Lb})
remaining, the critical behavior at the CSB infrared fixed point turns out to be identical to  
the ultraviolet asymptotical regime of the  
$2+1$-dimensional Gross-Neveu model which is renormalizable in the $1/N$-expansion,
$N$ being the number of Dirac fermion species \cite{HY}. 

Namely, the mean field (or $N=\infty$) scaling behavior of the boson propagator $<\phi\phi>
\propto 1/k$ gives rise to the
logarithmic divergence of the momentum integrals in Eqs.(\ref{G},\ref{Pi})
describing the first order $1/N$ corrections to the fermion and boson
wave functions as well as the gap and vertex functions.  
By differentiating these logarithmic corrections 
with respect to the external momenta,
one arrives at the usual renormalization group equations whose 
solution yields the infrared asymptotics of the renormalization factors    
\bea
Z_{p}&=&\left({p\over \Omega}\right)^{-\frac{2}{3\pi^2N}}\,\,\,\, ,\hspace*{5mm} 
\Delta_p=M\left({p\over \Omega}\right)^{\frac{2}{\pi^2N}} 
\label{Zfactor}
\eea
$$
\Pi_k-\Pi_0=k\left({k\over \Omega}\right)^{\frac{16}{3\pi^2N}}\,\, ,\hspace*{3mm}   
\Lambda_{p,k}=g\left({max(p,k)\over \Omega}\right)^{\frac{2}{\pi^2N}}
$$
from which the dimensions 
of the fermion $[\Psi]$ and boson $[\phi]$ fields can be readily read off.
Notably, the anomalous dimension of the fermion operator $[\Psi]-1\approx
1/3\pi^2N$ remains rather small even for $N=2$.

As regards the Lorentz-non-invariance of the bare
fermion Lagrangian (\ref{Lf}), in the quantum-critical regime 
the momentum integrals in Eqs.(\ref{G},\ref{Pi}) are dominated by 
the momenta parallel to the external ones, 
and, therefore, the anisotropic velocity factors can be scaled away 
without affecting the above power counting.

The critical exponents characterizing the CSB 
transition satisfy the hyperscaling relations which allow one to 
express all of them in terms of only two independent ones, e.g., 
the anomalous dimension exponent  
$\eta=2-D+2[\phi]\approx 1-(16/3\pi^2N)$ and the exponent  
$\nu=\beta/[\phi]\approx 1+(8/3\pi^2N)$ controlling the  
correlation length: $1/L_{\xi}=\delta m^2/(g^2N)\propto |g-g_c|^{\nu}$.
The latter remains finite in the symmetry broken 
(ordered) phase, as breaking discrete chiral symmetry does not result
in the appearance of a Goldstone mode below $T_c^*$. 

The fact that the critical exponents demonstrate
an explicit $N$-dependence implies that the universality class of the CSB 
transition is different for different $N$, the $D=3$ Ising  
transition ($\nu =0.63,~~\eta=0.025$) being recovered only in the limit $N\to 0$.

In the physical case of $N=2$, the above first order estimates for the 
critical exponents ($\nu\approx 1.14,~~\eta\approx 0.73$) compare favorably with 
the available Monte Carlo results ($\nu=1.00,~~\eta=0.75$) \cite{MC},
since the actual (inverse) parameter of  
this expansion $N{\rm Tr}({\bf 1}\otimes{\bf 1})=4N$ remains fairly large even for $N=2$.
For comparison, the first-order estimate for the anomalous dimension 
exponent given by the $\epsilon$-expansion of Ref.\cite{Sachdev} 
is $\eta\approx 4\epsilon/7=0.57$, although the agreement can be 
improved in the higher orders \cite{HY}.  
 
We emphasize that the above discussion pertains to the bulk properties of the  
layered $d$-wave superconductors and is, therefore, unrelated to 
the previously proposed scenario of a surface-induced $is$ pairing \cite{Greene}.

The alternate case of $id_{xy}$ pairing corresponds to 
a different choice of the coupling matrix ${\bf \Lambda}={\bf 1}\otimes
\sigma_3$ in Eq.(\ref{Lg}). This yields a chiral invariant order parameter 
${\overline \Psi_\alpha}{\bf 1}\otimes {\sigma_3}\Psi_\alpha=
\psi_{1\alpha}^\dagger{\sigma_2}\psi_{1\alpha}-\psi_{2\alpha}^\dagger
{\sigma_2}\psi_{2\alpha}$ which causes the term (\ref{Lg}) 
to explicitly break the chiral symmetry (\ref{Z2}). 

However, contrary to the $is$ order parameter, the $id_{xy}$ pairing and the corresponding
boson field $\phi$ change their signs under parity transformation ($x\to -x, y\to y$)
which acts upon the four-component spinors as  
$\Psi_\alpha\to \sigma_1\otimes\sigma_1\Psi_\alpha$.
It is this very property that allows for the existence of 
a zero-field spin or thermal Hall current in the $id_{xy}$-paired state \cite{Marston}. 

Therefore, by substituting parity for the chiral transformation (\ref{Z2}), one 
finds yet another $Z_2$ symmetry which gets spontaneously broken
upon the onset of the $id_{xy}$ order described by the same gap equation 
(\ref{gapeq}). The resulting critical behavior 
then appears to be identical to that found in the case of the $is$ pairing.
It will be, however, different from
the previously discussed second order $d\to d+id^\prime$ transition induced by
magnetic impurities \cite{Balatsky} as well as the first order one 
driven by external magnetic field \cite{Laughlin} (both these scenarios  
have recently been revisited and critically assessed in Ref.\cite{Woelfle}).

After having fully identified the nature of the second pairing transition,
we now proceed with computing the experimentally measurable
quasiparticle damping.
Whether attainable by varying a single physical parameter such as
doping or not, a zero temperature quantum critical point
affects the finite temperature dynamics of the system in a whole domain
of the $g-T$ plane where the correlation length $L_\xi$  
exceeds the thermal one $\propto 1/T$ (the dynamical exponent equals unity
due to the Lorentz invariance). 

In this quantum-critical regime,  
the renormalization cut-off is set by the temperature, and, 
therefore, the small energy- and
momentum-dependencies of the fermion and boson propagators are  
no longer governed by the anomalous operator dimensions from Eq.(\ref{Zfactor}).
Instead, the latter appear in the residues 
$\propto T^{2([\Psi,\phi]-1)}$ of these (now, pole-like) propagators.

After factoring out the fermion wave function renormalization factor,
the damping of the nodal quasiparticles with ${\bf p}=0$    
is given by the expression: $\Sigma(\varepsilon)={\rm Im}\{{\rm Tr}\,
[\,{\Gamma}_{0}(Z_p\langle\Psi{\overline \Psi}\rangle)^{-1}]\}$.
  
The universal function $\Sigma/T=\Phi(x,y)$ incorporates, alongside with the 
low- versus high-energy asymptotics for  
$x=\varepsilon/T <1$ ($>1$), the crossovers from the
quantum-critical to the other, 
renormalized classical and quantum disordered, regimes for $y=
\Omega(|g-g_c|/g_c)^\nu/T <1$ ($>1$), respectively.

The non-linear equation determining the 
function $\Phi(x,y)$ results from including $\Sigma(\varepsilon)$ into the 
fermion propagator in Eq.(6)  
\bea
\Sigma(\varepsilon)=\int{d\omega\over 2\pi}
\int \!{d{\bf q}\over (2\pi)^2}\left\{\tanh\frac{\varepsilon+\omega}{2 T}-
\coth\frac{\omega}{2 T}\right\}\nonumber\\
{\rm Im}\!\left[{\varepsilon+\omega+i\Sigma(\varepsilon+\omega) \over
(\varepsilon+\omega+i\Sigma(\varepsilon+\omega))^2-{\bf q}^2}\right]
\!{\rm Im}\!\left[{g^2 \over
{\Pi(\omega,{\bf q}) + m^2}}\right]
\label{gamma}
\eea
and, correspondingly, into Eq.(7)
for the spatially isotropic, albeit manifestly non-Lorentz-invariant, 
finite temperature polarization operator 
$$
\Pi(\omega,{\bf q})={g^2N} 
\int{d\varepsilon\over 2\pi}
\int \!{d{\bf p}\over (2\pi)^2}
\left\{\tanh\frac{\varepsilon+\omega}{2 T}-
\tanh\frac{\varepsilon}{2 T}\right\}
$$
\be
{{(\varepsilon +i\Sigma(\varepsilon))
(\varepsilon+\omega+i\Sigma(\varepsilon+\omega))-{\bf p}({\bf p}+{\bf q})} 
\over
{[(\varepsilon+i\Sigma(\varepsilon))^2-{\bf p}^2]
[(\varepsilon+\omega+i\Sigma(\varepsilon+\omega))^2-({\bf p}+{\bf q})^2]}}
\label{Pi_T}
\ee
Notably, in the quantum-critical regime ($y < 1$ ) the 
energy-independent renormalization factors drop out from Eqs.(\ref{gamma},\ref{Pi_T}), 
thanks to the underlying Ward identities of the HY model \cite{HY}.

Both analytical and numerical, analyses of the coupled equations 
(\ref{gamma},\ref{Pi_T}) show that   
the fermion damping behaves as $\Sigma\propto T$ and $\propto\varepsilon$ 
at high and low temperatures, respectively, in agreement with experiment \cite{Valla}
and the results of Ref.\cite{Sachdev}.

Below the crossover ( $y>1$ ) to the quantum disordered (weak coupling, low-$T$)
regime, the infrared cut-off is provided by the (inverse) correlation length $L_\xi^{-1}$, and 
the damping of the Lorentz-invariant nodal quasiparticles appears to retain
its energy dependence even for $\varepsilon < T$, unlike the case of 
fermions with extended Fermi surface.  
In the energy intervals $x\geq 1$, $1/y^4<x<1$, and $x\leq 1/y^4$, 
the self-consistent solution of Eqs.(9,10) behaves as 
$\Phi(x,y)\propto x^3/y^2$, $\propto x^{1/2}/y^2$, and $\propto 1/y^4$, respectively.
For $\varepsilon\sim T$ it reproduces the perturbative second order 
result $\Sigma\propto T^3$ (see \cite{Scalapino}) 
valid for a generic short-ranged coupling, while 
for $\varepsilon\to 0$ we obtain $\Sigma\propto T^5$, 
one possible experimental implication
being a significant narrowing of the nodal ARPES linewidth below the crossover.

Likewise, in the renormalized classical (strong coupling, low-$T$) regime one can expect a  
suppression of the nodal quasiparticle density of states
that can manifest itself in tunneling and specific heat data
as well as a sharp decrease of both, thermal and optical,   
conductivities accompanied by an increasing thermal Hall angle \cite{Ong}.

Furthermore, according to experiment, the linear temperature/energy dependence of the
quasiparticle damping extends well into the normal phase \cite{Valla}. This might be  
suggestive of a possibility of applying the HY model to the pseudogap state, 
as the above discussion relies on the presence of the $d_{x^2-y^2}$-symmetrical (pseudo)gap 
in the local quasiparticle spectrum rather than the onset of 
the superconducting coherence across the entire system.
To this end, it should be interesting to contrast the quantum critical
scenario described in this paper
against the results of the previous approaches to the pseudogap problem
which have focussed on the couplings between the nodal Dirac fermions
and magnetic order parameters \cite{Fisher} as
well as classical \cite{Millis} and quantum \cite{Wen} superconducting phase fluctuations.

To summarize, 
we carried out a non-perturbative analysis
of the secondary pairing transition in planar $d$-wave superconductors.
By identifying the transition in question with the HY model of the 
$2+1$-dimensional Dirac fermions we succeeded in finding all the critical exponents
and determining the behavior of the quasiparticle propagators. 
At $\varepsilon > T$ the quantum-critical fermion propagator 
was found to have a numerically small anomalous
dimension, while in the opposite limit it exhibits
linear damping, as observed in experiment. 

Should a secondary pairing occur in the bulk 
$d$-wave superconductors upon tuning the phenomenological
parameter $g$ in a (yet, unspecified) way, it can manifest itself 
as a marked change in quasiparticle spectral and transport properties even
in the absence of boundaries (cf.\cite{Greene}), magnetic impurities (cf.\cite{Balatsky}), 
and/or external magnetic field (cf.\cite{Laughlin}).
Also, the two-dimensional spatial parity's remaining intact or being spontaneously broken  
may allow for an experimental discrimination 
between the two different kinds ($is$ versus $id_{xy}$) of incipient ordering  
with broken time-reversal symmetry. 

This research was supported by the NSF under Grant No. DMR-0071362 (DVK).

\wt
\end{document}